\documentstyle[12pt]{article} 
\begin{document} 
 
\def\bce{\begin{center}} 
\def\ece{\end{center}} 
\def\beq{\begin{eqnarray}} 
\def\eeq{\end{eqnarray}} 
\def\ben{\begin{enumerate}} 
\def\een{\end{enumerate}} 
\def\ul{\underline} 
\def\ni{\noindent} 
\def\nn{\nonumber} 
\def\bs{\bigskip} 
\def\ms{\medskip} 
\def\tr{\mbox{tr} } 
\def\wt{\widetilde} 
\def\wh{\widehat} 
\def\brr{\begin{array}} 
\def\err{\end{array}} 
\def\dsp{\displaystyle} 
\def\eg{{\it e.g.}} 
\def\ie{{\it i.e.}} 
 
\hfill IEEC/CSM-99-62 
 
\hfill hep-lat/9906038 
 
\hfill July 1999 \, (revised)
 
\thispagestyle{empty} 
 
\vspace*{15mm}

\begin{center} 
 
{\LARGE \bf  On the continuum limit of the lattice 
chiral anomaly trace relation}

\vspace{12mm} 
 
\medskip

{\sc E. Elizalde}\footnote{\ni E-mail: 
elizalde@ieec.fcr.es \  eli@ecm.ub.es \ \ 
http://www.ieec.fcr.es/cosmo-www/eli.html}\\ 
Consejo Superior de Investigaciones Cient\'{\i}ficas (CSIC),\\ 
Institut d'Estudis Espacials de Catalunya (IEEC), \\ 
Edifici Nexus 201, Gran Capit\`a 2-4, 08034 Barcelona, Spain\\ and \\ 
Departament ECM i IFAE, Facultat de F\'{\i}sica, \\ 
Universitat de Barcelona, Diagonal 647, 
08028 Barcelona, Spain \\

\vspace{20mm} 
 
{\bf Abstract} 
 
\end{center} 
 
Different aspects concerning the rigorous definition of the traces and  
determinants of the operators involved in a procedure 
---proposed by Neuberger and others--- for avoiding  
fermion doublers on the lattice, are considered. 
A result of the analysis is that it seems unclear that the  
consequences derived from the independent treatment of the traces of the  
two operators contributing to the index relation on the lattice, 
as carried out in recent manuscripts, can be given rigorous  
mathematical footing, 
in particular, that these treatments can commute with the continuum limit 
---the lattice regularized trace being additive all the way through 
the limit, while the 
otherwise regularized trace in the continuum is not so.

\vfill 
 

\newpage 
 
As emphasized elsewhere \cite{p1},  
many fundamental calculations of Quantum Field Theory reduce, in essence,  
to the computation of determinants and traces  of operators.  
Important as the concept of determinant of a differential (or  
pseudodifferential) operator may be 
for theoretical physicists \cite{elicmp,eli2}, it is  
surprising that this seems not to be a subject of study among 
function analysts or mathematicians in general. As a consequence,  
theoretical physicists are often left with the burden of having to give 
sense  to determinants that involve in its definition 
some kind of regularization.  The subject 
has many things in common with that of divergent series but has not been 
so groundly investigated and lacks any reference comparable, \eg, to the  
beautiful book by Hardy \cite{hardy}. 
There are  well stablished theories of determinants for 
degenerate operators, for traceclass operators in the Hilbert space, 
 Fredholm operators, etc. \cite{kato} 
but, again, these definitions of determinant do not fulfill all the needs  
 which arise in QFT. Try to answer, from the books, questions like: 
 What is the value of the determinant of minus the identity operator 
in an infinite dimensional space? And that of the spectral product 
$\prod_{n \in N} 
(-1)^n$? Is the last actually equal to the product of the separate
determinants of 
the plus 1s and of the minus 1s? A detailed disquisition on this situation, 
with a number of basic examples and possible uses in Physics, has appeared 
recently \cite{p1}. 
 
Here I  will concentrate on issues related with an important 
development that is taking place in the treatment of fermions in lattice 
gauge theory \cite{neub1}. The literature  is so extense and dynamic  
that it is almost impossible to keep track of all the relevant contributions, 
which can,  
however, be recovered substantially from the references in the papers 
I will mainly concentrate on: two recent preprints by H. Neuberger 
\cite{neub2} and K. Fujikawa 
\cite{fuji1}.  
 
There is general agreement that the starting  point of this development  
is the use of the Hermitian lattice Dirac operator $\gamma_5D$ satisfying 
the Ginsparg-Wilson relation \cite{GW} 
\beq 
\gamma_5D + D\gamma_5 = a D\gamma_5D,  \label{GW1} 
\eeq 
with $\gamma_5$ a Hermitian chiral Dirac matrix. 
The important contribution of Neuberger and Narayanan \cite{neub1,neub2} 
(inspired in work by Kaplan \cite{K} and Frolov and Slavnov \cite{FS}), 
has been to construct  
an explicit operator satisfying this relation and free from fermionic 
species doubling. 
As specified  in \cite{neub2}, this operator ---the massless fermionic matrix  
$D$--- is given by 
\beq 
D=\frac{1}{a}(1+\epsilon'\epsilon), 
\eeq 
where $\epsilon$ and $\epsilon'$ are Hermitian and square to unity, so that  
$V=\epsilon'\epsilon$ is unitary. By the way, it is also said in 
\cite{neub2}  
that, det $\epsilon = (-1)^{\frac{1}{2}\tr \, \epsilon}$, and that the same 
yields  
for $\epsilon'$. 
This statement is certainly true for an operator  $\epsilon$ acting in a
finite-dimensional space,\footnote{I am indebted with H. Neuberger for a 
critical and illuminating remark on this point, that led me to rectify a
statement in a previous 
version of this manuscript.} since it also acts on spinorial indices of 
four-dimensional Dirac spinors, so that its dimension is divisible by four
 (and finite). Calling $n_\pm$ be the number of $\pm 1$ eigenvalues of 
$\epsilon$, respectively, it turns out that  
\beq
\det \epsilon = (-1)^{-n_-} = (-1)^{-n_- + (n_- +n_+)/2} =
    (-1)^{\frac{1}{2} \tr \epsilon}. \label{n1}
\eeq

Let us observe that for a Hermitian operator $H$ squareing to unity in an 
infinite-dimensional space we would have generically the following (working in 
diagonalized form):
\beq
H^2 =I=\mbox{diag } \left( e^{2\pi i k_j} \right)_{j\in J}, \qquad  
\log H = \mbox{diag } \left( \pi i k_j \right)_{j\in J}, \quad k_j\in Z, 
\eeq
with $J$ a countable set, from where
\beq
\log \det H= \tr \log H = \pi i \sum_{j\in J} k_j, \qquad  \det H= 
e^{\pi i \sum_{j\in J} k_j} = (-1)^{n_-},
\eeq
since $k_j$ can be fixed to be 0 for the $j$ corresponding to positive 
eigenvalues of $H$, and 1 for the negative. The very last equality
 is far from naive.  The sum here is an infinite series, however $n_-$ 
is not infinite, but the finite number obtained from a 
regularization of the determinant that respects the $\log \det = \tr \log$
condition (see Ref. \cite{p1} complete for details and examples). The argument 
leading to Eq. (\ref{n1}) would not hold in general in this infinite 
dimensional case. However, for a  unitary operator, $U$, under the  
assumption that it squares to minus unity, $U^2=-I$, we would have instead
\beq 
2\log U= \mbox{diag } \left( \pi i (\pm 1+ 2 k_j) \right)_{j\in J},
\quad k_j\in Z, 
\eeq 
and {\it all} $k_j$ could be fixed to be 0 now, therefore, 
\beq 
 \det U = e^{\tr \, \log U}=e^{\frac{\pi}{2}\tr \, U}= i^{n_+-n_-}.
 \label{tdV1} 
\eeq 
In this case, when going to the infinite dimensional limit by use of any 
apropriate regularization (just respecting the rule  $\log \det = \tr \log$),
and being now $n_+$ and $n_-$ the finite values assigned by this 
regularization, we are sure to preserve the index relation in a natural way,
all along the regularization process. The simplest example of such an
operator $U$ is obviously
$U=i\epsilon$. That $V$ cannot be such operator is clear from the observations
of Neuberger in \cite{nplb}. Thus, we see that in order to preserve 
the index relation in the infinite  limit, through any reasonable 
regularization process (\ie, respecting the trace-log identity)
one just needs to complexify the operator $\epsilon$ in a simple manner.
 
The operator $D$ above, introduced by Neuberger ---and which  
satisfies the Ginsparg-Wilson relation (\ref{GW1})--- can be written as  
follows  
\beq 
D= \frac{1}{a} \left( 1- \gamma_5  \frac{H_W}{\sqrt{H_W^2}} \right) = 
\frac{1}{a} \left( 1+   \frac{D_W}{\sqrt{D_W^\dagger D_W}} \right), 
\eeq 
where $D_W$ is the Wilson lattice Dirac operator. In the  
decomposition $V=\epsilon'\epsilon$ above, $\epsilon'$ corresponds to  
$\gamma_5$, while $\epsilon$ is given by the sign function  on the  
Hermitian operator $H_W$, that is 
\beq 
\epsilon (H_W) = \frac{H_W}{\sqrt{H_W^2}}. 
\eeq 
As a consequence, $\epsilon$ is defined for all gauge orbits with  
$H_W^2>0$, being nonanalytic when $H_W$ has a zero eigenvalue. The exclusion 
of the set of gauge fields where $H_W$ develops zero modes is necessary 
in any lattice calculation, in order to partition 
the space of lattice gauge orbits into the different topological sectors. 
The concept of measure in a finite lattice is mathematically well settled,
as the product of Haar measures for each link times the exponent of the
pure gauge action (assumed $C^\infty$ and bounded in the link
variables) and the fermionic determinant.\footnote{Thanks are given to H. 
Neuberger for such a precise definition.}
But the fine question still arises: what would be the  {\it measure} of 
this set of gauge fields in an {\it infinite} lattice? The concept of a zero 
measure  set (and of measure of a measurable set in general) is  
a perfectly stablished one in  Measure Theory, with its imprints in  
Functional Analysis and, in particular, in the Theory of Hilbert Spaces. 
However, when a regularization process  
is involved in QFT,  concepts as this one ---and  as those of trace or  
determinant (such as in Eq. (\ref{tdV1}))--- loose their ordinary meaning 
(together with some of their fundamental properties) on 
being submitted to the regularization process \cite{p1}. For instance, 
as is carefully explained in \cite{p1}, in the calculation of a trace 
or a determinant through the zeta function (which is, by the way, one 
of the most rigorous definitions of regularization available), the  
would be ``measure'' of the set of natural numbers is (after regularization), 
equal to $-1/2$. This obviously contravenes already the first axiom of the  
definition of measure: aside from being negative, it is actually smaller  
than the ``measures'' of the individual components of the set (each being 
equal to  1). And this is only an example, the most simple one available.  
Under such perspective, how can we really know that the  
contribution of the zero mesure set above, after regularization, will
be insignificant in an infinite lattice, as compared to the regularized 
contribution of the rest? 
Indeed, one may argue that the lattice itself provides already {\it the} 
regularization in lattice theory, 
but then the question can be immediately translated to the continuum limit, 
which might not 
commute with  some of these calculations carried out on the lattice. 
I think this is not an easy question to answer in general and can only be 
solved, in principle, through specific models.  We will come back to this 
point in what follows.  Notice, on the other hand, that this kind of 
considerations, in particular, that $c+c+c+\cdots =\zeta (0) \, c =-c/2$, 
are essential to the Frolov-Slavnov's generalized Pauli-Villars 
regularization \cite{FS} with an infinite number of Pauli-Villars fields. 
As clearly explained in \cite{neubv}, one of the successes of 
the overlap construction was that, having been obtained by infinite iteration
from an explicit chiral starting point, aside incorporating non-perturbative 
contributions it could be recast in a way that made in the end
no reference to anything infinite (in particular, to the infinite fermions
introduced in the process). In principle, however, 
the continuum limit of the lattice regularization remains to be taken.
This issue will be considered in what follows.   
 
Not abandoning the regularization scheme, an elusive argument affecting 
the definition of the trace of the operators involved in the above theory will 
be now rised. An important and very direct consequence of the 
Ginsparg-Wilson relation (\ref{GW1}) is that it has allowed for the  
introduction of an index on the lattice, first defined in the
overlap literature as given by $\tr \, \epsilon /2$. The index relation  
can be written as \cite{irl} 
\beq 
\tr \, \left[\gamma_5 \left( 1- \frac{a}{2}D \right) \right] = n_+-n_-, 
\label{irl1} 
\eeq 
being $n_\pm$ the number of normalizable zero modes from 
\beq 
\gamma_5  D \phi_n =\lambda_n \phi_n 
\eeq 
simultaneously satisfying the eigenvalue equations  
\beq 
\gamma_5  \phi_n =\pm \phi_n, 
\eeq 
respectively. Fujikawa \cite{fuji2}  has a beautiful proof of this relation 
that does not use at all that tr $\gamma_5=0$, a relation which is  
supposed to hold in fact in lattice theory, in contradistinction to the  
chiral anomaly relation in the continuum, namely 
\beq 
\tr \, \gamma_5 =n_+-n_-,  \label{11} 
\eeq 
that plays a fundamental role in the path-integral treatment of the anomaly 
\cite{it2}. 
As observed by Chiu, the traceless condition for $\gamma_5$  
on the lattice poses the constraint \cite{chiu1}: 
\beq 
\tr \, \gamma_5 =n_+-n_-+N_+-N_-=0,   \label{12} 
\eeq 
where the $N_\pm$ correspond to the number of eigenstates of  
$\gamma_5 D  \phi_n =\pm (2/a)  \phi_n$, satisfying also  
$\gamma_5 \phi_n =\pm  \phi_n$, respectively. Fujikawa proves in \cite{fuji2} 
that, for the operator $D$, the relations  
\beq 
\tr \, \gamma_5 =n_+-n_-,\qquad  \tr\left[ -(a/2) \gamma_5D\right] =0, 
\label{13} 
\eeq 
are actually consistent on the lattice,  
in agreement also with the fact that species doublers  
are missing for the operator $D$. It is recognized in  \cite{fuji2}, that  
the evaluation of  $\tr \, \gamma_5$  is somewhat subtle, and also that it is  
important to observe that  the relation $\tr \, \gamma_5=0$ must hold for 
any sensible basis in a given theory that may be used to define the trace. 
 
Now, as  it turns out, the fermionic operators one is dealing with here 
in the continuum are {\it not} of the trace class. Moreover,   
as is clearly explained in \cite{p1}, for, \eg , the zeta  
function definition of the regularized trace a multiplicative anomaly of the 
determinant exist, that is intimately  related with the fact that the  
regularized 
trace does {\it not} satisfy, in general, the additive property, that is,  
generically, 
\beq 
{\tr}_\zeta (A+B) \neq {\tr}_\zeta A + {\tr}_\zeta B.   \label{14} 
\eeq 
This results immediately in the appearance of the  
multiplicative anomaly of the determinant, namely the fact that one 
also has in general \cite{p1,ma1}: 
\beq 
{\det}_\zeta (A B) \neq ({\det}_\zeta A) \ ({\det}_\zeta B),   \label{14d} 
\eeq 
even for commuting operators. 
The absence of the additive property for the regularized trace has led to 
a considerable number of errors in the past. Here, it represents a serious 
drawback to the rigor of the derivation, should the additivity be taken 
for granted. 
 
In fact, in all the calculations being carried out on the lattice, it is 
obviously true that the additive property of the trace is preserved 
(the reason being that one {\it always} works with a {\it finite} lattice). 
In particular, for the traces in  (\ref{11})--(\ref{13}), relevant for 
performing the continuum limit of the lattice index relation, one always 
uses \cite{fuji1,fuji2,chiu1,chiu2} 
\beq 
\tr \, \left[ \gamma_5 \left( 1-\frac{a}{2} D \right) \right] = 
\tr \, \gamma_5 - \tr \, \left(\frac{a}{2} \gamma_5 D \right). 
\label{tt} 
\eeq 
This is obviously true for any {\it finite} lattice which, on the other hand, 
is the kind always employed for performing the numerical computation. 
 
One should notice, however, that the breaking  (\ref{14}) of the trace additive 
property may happen already for a {\it discrete} infinite space and for 
extremely simple operators, as the identity itself. In fact, 
for instance, for the operators 
\beq  
O_1 = \mbox{diag\ } (1,2,3,4, \ldots ), \qquad  
O_2 = \mbox{diag\ } (1,1,1,1, \ldots ) \equiv  I,  
\eeq  
and their sum  
\beq  
O_1 +O_2 = \mbox{diag\ } (2,3,4,5, \ldots ),  
\eeq  
we have that the corresponding $\zeta$-traces are:  
\beq  
&&\tr_\zeta O_1 =  \zeta_R (-1) = - \frac{1}{12},  \qquad  
\tr_\zeta O_2 =  \zeta_R (0) = - \frac{1}{2},  \nn \\ &&  
\tr_\zeta (O_1 +O_2)=  \zeta_R (-1) -1= - \frac{13}{12},  
\eeq  
the last trace having been calculated according to the rules of  
infinite series summation (see \eg, Hardy \cite{hardy}). We observe that  
\beq  
\tr_\zeta (O_1 +O_2) - \tr_\zeta O_1 - \tr_\zeta O_2=  - \frac{1}{2} \neq 0 
\qquad !! 
\eeq  
If this happens in such  simple situation, involving the identity  
operator, one can easily imagine that any precaution one can take in  
manipulating infinite sums might  
turn out to be insufficient. In other words, generically, the additive 
property for the regularized trace in the continuum is broken, and so is 
it also for a discrete {\it infinite} lattice. 
 
Now, when one performs the continuum limit of the lattice ---involving at 
every step a {\it finite} lattice--- the trace additive property is preserved, 
all the way through the  limit. But we know that this property is broken 
in fact in the continuum limit, by {\it any} of the usual regularizations 
(as zeta or Pauli-Villars). This is, in our view, an inconsistency that 
renders to the level of `formal' 
many proofs of the continuum limit of the celebrated index 
relation on the lattice, Eq. (\ref{irl1}) \cite{irl}, giving as a result 
the very well known index theorem in the continuum QFT \cite{abp}. 
 
Summing up, there is generically an inconsistency in calculating the 
two traces that appear on the rhs of Eq. (\ref{tt}) {\it separately} 
and  pretending that the continuum limit of their sum will coincide with 
the trace of the continuous anomaly operator, as a whole. An additional 
contribution is generically missed in the process \cite{p1}. Now, the 
point is in fact  extremely subtle, for the usual argument is that, 
all the way to the limit, the equality in Eq. (\ref{tt}) holds, 
this being, of course, because the 
infinite continuum is replaced in the lattice regularization by a  finite
discrete lattice. If an infinite discrete lattice were used, the 
trace additive property would in general get lost already ---as happens 
with other regularizations of the continuum QFT theory (as zeta or 
Pauli-Villars)--- and this would occur already {\it before} performing 
the continuum limit.  Moreover, there is also an elusive point concerning 
the question of the zero-measure of the set of null eigenfunctions of 
Neuberger's $D$-operator when taking this limit. I envisage a similar 
difficulty to the above one of the trace there: it would show up as soon 
as we considered an infinite lattice. 
 
As a conclusion, on top of the inconsistency discovered by Fujikawa in 
performing the limit $a\rightarrow 0$ (continuum limit) in lattice gauge 
theory when one starts from the condition $\tr \, \gamma_5 =0$, we have 
pointed out here a potential additional flaw, that 
precludes the independent limits of the two contributions to the trace 
defining the anomaly from having sense individually. 
The fact that the correct continuum limit for the chiral anomaly in the 
continuum QFT has been obtained starting from the chiral anomaly 
relation on the lattice ---and this using a number of quite different 
procedures (ocasionally fine-tuning ones, see, \eg, Chiu \cite{chiu2}, 
Table 2, where precise numerical comparisons with the exact result on the 
torus \cite{sw1} are given)--- does not imply that there is a 
rigorous proof of the continuum limit. On the contrary, our arguments 
in the present paper have uncovered a possible mathematical inconsistency 
that points towards the conclusion that the demonstrations that involve 
an unallowed (in the continuum) trace splitting, might 
remain valid at the formal level only. Therefore, the problem would still 
persist of going from the condition 
$\tr \, \gamma_5 =0$ on the lattice ---as demanded specifically by the 
index relation, a condition that following Fujikawa is actually 
inconsistent with the absence of non-physical fermion doublers 
on the lattice--- to the condition $\tr \, \gamma_5 =n_+-n_-$ in the continuum 
(that follows from the path integral derivation of the index 
theorem \cite{it2} for a chiral QFT), in a smooth and mathematically 
rigorous way. 
 
\bs 
 
 
\vspace{3mm} 
 
\noindent{\bf Acknowledgments} 
 
Thanks are given to H. Neuberger for important correspondence.  
This investigation has been supported by DGICYT (Spain), project 
PB96-0925, by CIRIT (Generalitat de Catalunya), by the Italian-Spanish 
program INFN--CICYT, and by the 
German-Spanish program Acciones Integradas, project HA1997-0053. 
 

\end{document}